# Exotic rare earth-based materials for emerging spintronic technology

Sachin Gupta

Department of Physics, Bennett University, Greater Noida 201310, India

Email: sachin.gupta@bennett.edu.in; gsachin55@gmail.com

## Abstract

The progress in materials science has always been associated with the development of functional materials systems, which enables us to design proof-of-concept devices. To advance further, theoretical predictions of new novel materials and their experimental realization is very important. This chapter reviews the intriguing properties of rare earth-based materials and their applications in spintronics. Spintronics is an emerging technology, which exploits spin degree of freedom of an electron along with its charge property. Discovery of various physical phenomena and their industrial applications in the field of magnetic sensors, magnetic recording and non-volatile memories such as magnetic random access memory (MRAM) and spin-transfer torque (STT) MRAM opens several new directions in this field. Materials with large spin polarization, strong spin-orbit coupling, and tunable electronic and magnetic properties offer an excellent platform for the spintronics technology. Combination of rare earths with other elements such as transition metals show broad range of structural, electronic, and magnetic properties which make them excellent candidates for various spintronic applications. This chapter discusses many such materials ranging from Heusler alloys, topological insulators to two-dimensional ferromagnets and their potential applications. The review gives an insight of how rare-earth materials can play a key role in emerging future technology and have great potential in many new spintronic related applications.

**Abbreviations**

| | |
|---|---|
| 2D | Two-dimensional |
| 2DM | Two-dimensional magnet |
| AFM | Antiferromagnetic |
| ARPES | Angle resolved photoemission spectroscopy |
| DFT | Density functional theory |
| DoF | Degree of Freedom |
| DoS | Density of states |
| DW | Domain wall |
| EQHA | Equiatomic quaternary Heusler alloy |
| FCF | Fully compensated ferrimagnet |
| FM | Ferromagnetic |
| GMR | Giant magnetoresistance |
| GGG | Gadolinium gallium garnets |
| GSGG | Gadolinium scandium gallium garnets |
| HA | Heusler Alloy |
| HM | Half metallic |
| HDD | Hard disk drive |
| HRXRD | High-resolution X-ray diffraction |
| ISHE | Inverse spin Hall effect |
| MBE | Molecular beam epitaxy |
| ML | Monolayer |
| MTJ | Magnetic tunnel junction |
| MTI | Magnetic topological insulator |
| MR | Magnetoresistance |
| MRAM | Magnetic random-access memory |
| PM | Paramagnetic |
| PMA | Perpendicular magnetic anisotropic |
| QAHE | Quantum Anomalous Hall effect |
| RE | Rare earth |
| REIG | Rare earth iron garnet |
| SdH | Shubnikov-de Haas |
| SOC | Spin-orbit coupling |
| SOT | Spin orbit torque |
| STT | Spin transfer torque |
| SGGG | Substituted gadolinium gallium garnets |
| SQUID | Superconducting quantum interference device |
| TDS | Topological Dirac semimetal |
| TI | Topological Insulator |
| TIG | Thulium iron garnet |
| TM | Transition metal |
| TMR | Tunnel magnetoresistance |
| TRS | Time reversal symmetry |
| TSS | Topological surface states |
| TWS | Topological Weyl semimetal |
| VSM | Vibrating sample magnetometer |
| YIG | Yttrium iron garnet |

## 1. Introduction

Rare earths (REs) consist of elements from lanthanide series (La-Lu, Ln) and including Sc and Y. Despite their name they are relatively in ample amount in earth's crust. REs in combination with other elements have been studied for many years and are still topic of great interest to many researchers because of their intriguing physical properties and application potentials [1,2]. By employing REs, one can benefit from properties such as large magnetic moments, strong spin-orbit coupling (SOC) or magnetocrystalline anisotropy due to their special electronic configuration. The general electronic configuration of REs (excluding Y and Sc) is $[Xe]4f^{0-14}5d^{0-1}6s^2$. It can be noted that 4f shell electrons in REs are well localized within the atom, which is shielded by $5s$ and $5p$ states from the surroundings. The magnetism in RE originates from the partially filled 4$f$ shell electrons. To a large extent, the magnetism in rare earth solids therefore resembles that of free rare earth atoms. Furthermore, localization of 4f electrons results in large magnetic moments. In case of ordered solids, REs offers the strongest ferromagnets, which therefore enables them for technological applications involving permanent magnets. The outer 5$d$ and 6$s$ electrons, on the other hand, are delocalized and behave as conduction electrons, which contribute very little to the magnetic moment but they play a role in mediating exchange interactions, and therefore in determining the magnetic properties in the solid [3,4]. Since the strength of SOC is proportional to $Z^4$, heavier atoms (Z > 56) in the REs results in strong SOC. When these REs combine with transition metals, these are complemented with the high magnetic transition temperatures. All these properties play important roles in the use of RE-based materials. In recent years, these materials have been employed in a range of applications such as permanent magnets, medical equipment, magnetic refrigeration and sensors [3,5–10]. In addition, RE materials also find a special place in spintronic applications due to their interesting physical properties. The chapter highlights various phenomena linked to spintronic technologies and their potential applications .

## 2. Spintronics

Spintronics is a rapidly growing multidisciplinary field, which utilizes spin degree of freedom (DoF) of an electron instead of its charge DoF. Exploitation of spin DoF of an electron or in other words spin polarized-transport leads to many advantages over conventional electronics such as faster data operation, low energy consumption, non-volatility of data storage, high storage density [11]. Discovery of various physical phenomena and their industrial applications in the field of magnetic sensors, magnetic recording and non-volatile memories such as magnetic random access memory (MRAM) and spin-transfer torque (STT) MRAM opens several new directions in this field [12]. One of such discoveries, which revolutionized the spintronic research is giant magnetoresistance (GMR) effect.

**2.1 Giant magnetoresistance (GMR)**

The change in the electrical resistance of a material on application of a magnetic field is referred to as magnetoresistance (MR) and is of great technological importance. The larger the magnitude of MR is, the better for technological applications as it increases the device stability. Therefore, the search of material systems with larger MR is very important for technological applications.

Grünberg *et al*., [13] and Fert *et al*. [14] in 1988 independently discovered GMR effect in multilayer system comprised of ferromagnetic (FM) and non-magnetic conductive layers. Figure 1 shows one of such multilayer system including FM Fe layers separated by thin non-magnetic Cr layers (few atomic planes) [14]. The thickness of the thin Cr layers plays a key role in implementing the exchange coupling (FM or antiferromagnetic (AFM)) between Fe layers. The electrical resistance changes significantly depending upon the magnetization direction of the two FM layers (Figure 1), which is determined by the exchange coupling between the adjacent FM layers. The electrical resistance is high for the AFM (antiparallel) exchange coupling between Fe layers. When a magnetic field is applied, the exchange coupling changes from AFM to FM (parallel) and results in drastic reduction in electrical resistance of the system. GMR effects of 79% at 4 K and 20% at 300 K have been observed in Fe-Cr multilayer systems at low magnetic fields [14]. The enormous change in the magnitude of the electrical resistance justifies the term 'Giant' in GMR effect. The change in the electrical resistance is attributed to the spin-dependent scattering of an electron in the multilayer system. In this way, one can have high and low electrical resistance, which can be read as '0' and '1' in binary language. After the demonstration of GMR effect, there was extensive research on the underlying physics of spin dependent transport using a variety of theoretical models, engineering magnetic systems and combining various materials [15]. Along with basic research, the application potential of the GMR effect was identified by magnetic recording media and GMR read heads were introduced in hard disk drive (HDD). In addition to HDD, the GMR effect has also been employed in magnetic sensors and MRAM technology. A typical configuration used for GMR effect is a spin-valve, which consists of two FM layers separated by a conducting non-magnetic layer. One of these FM layers is a hard magnetic layer (i.e. having a large coercive field), called fixed or pinned layer and the other is a soft magnetic layer (i.e. having a low coercive field), also called free layer. Hence upon application of a magnetic layer, the free layer can be easily aligned in the magnetic field direction while the pinned layer is insensitive to moderate magnetic fields. In this way, one can easily change the magnetization alignment of two FM layers and hence significantly change the electrical resistivity.

S. Gupta, Handbook on the Physics and Chemistry of Rare Earths, Vol. 63, (2023).

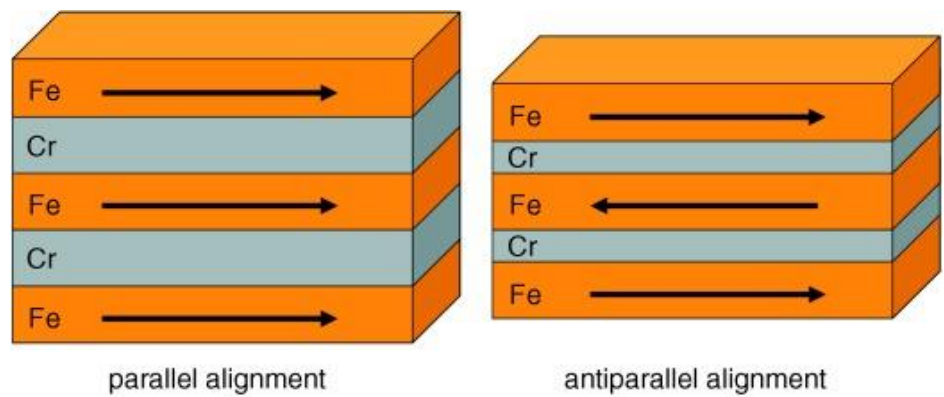

**Figure 1.** Schematic of a Fe/Cr multilayer system, showing giant magnetoresistance (GMR) effect. The two blocks of Fe-Cr multilayer system show ferromagnetic and antiferromagnetic coupling between adjacent Fe layers. Reproduced with permission from [11] © 2007, Wiley-VCH Verlag GmbH.

**2.2 Tunnel magnetoresistance (TMR)**

When the non-magnetic conductive layer between two FM layers is replaced by a thin insulating layer (usually $Al_2O_3$, MgO) the structure behaves as a tunnelling junction. In this type of multilayer system, electrons tunnel through thin insulating barriers and therefore the resultant magnetoresistance is called tunnel magnetoresistance (TMR). It is worth to note here that the physics of TMR is different from the GMR effect since tunnelling is involved in the TMR effect. The breakthrough in this field took place in 1995 when Moodera *et al.* [16] and Miyazaki *et al.* [17] demonstrated an unprecedented large TMR, which was much larger than the GMR in similar systems having a conductive space layer [15]. It was reported that TMR with an $Al_2O_3$ insulating layer is larger than the GMR and the effect was recently further enhanced with the use of an MgO insulating barrier. The larger effect introduces more stability in the devices and hence, presently, GMR effect in hard disk drive read heads and MRAM is replaced by TMR effect. Figure 2 shows a timeline of development of GMR and TMR in multilayer systems.

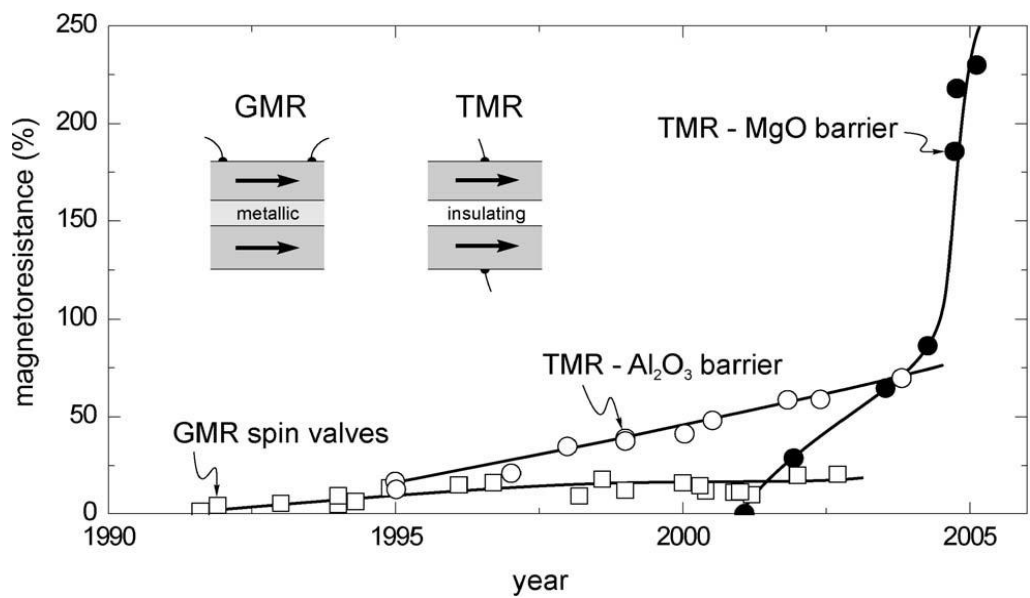

**Figure 2.** Timeline of development of giant magnetoresistance (GMR) and tunnel magnetoresistance (TMR) in multilayer systems at room temperature. GMR data are shown for spin valves having two ferromagnetic layers separated by a metallic space layer while TMR data are shown for tunnelling barriers of $Al_2O_3$ as well as MgO. Reproduced with permission from [15] © 2008 Elsevier B. V.

According to Jullière's model [18], if a magnetic tunnel junction (MTJ) has two electrodes with spin polarization $P_1$ and $P_2$, the TMR in MTJ can be estimated as

$$TMR = \frac{2P_1P_2}{1 - P_1P_2}$$

The spin polarization $P$ of a system is then described as

$$P = \frac{N_\uparrow - N_\downarrow}{N_\uparrow + N_\downarrow}$$

Where $N_\uparrow$ and $N_\downarrow$ are the density of majority and minority electrons at the Fermi level, respectively.

S. Gupta, Handbook on the Physics and Chemistry of Rare Earths, Vol. 63, (2023).

Ferromagnetic transition metals (in the form of alloys or compounds) are commonly used as magnetic electrodes in GMR/TMR devices. Due to their intrinsic magnetism and variation in the coercive fields, rare earth materials have also been tested in such devices. Recently, Warring *et al.* [19] fabricated MTJs using SmN and GdN as electrodes and observed a TMR of 200% in these MTJ.

## 3. Rare earth based spintronic materials

Exploitation of the spintronic technology depends on the development of new functional materials. Theoretical predictions and experimental realization of new materials help in engineering future spintronic devices. In this section, material aspects and basic physical principles, which play an important role in the progress of spintronics, are presented and discussed. Moreover, the discussion on materials highlights how rare earth-based materials can be an integral part of emerging spintronic technology.

### 3.1 Half-metallic materials

Half-metallic (HM) ferromagnetic materials have drawn hefty attention for spintronic applications due to their exceptional electronic structures [20,21]. Figure 3 shows a schematic of density of states (DoS), $n(E)$ as a function of energy, $E$ for HM materials, where two sub-bands show completely different nature. It can be noted that HM materials show metallic (conducting) behavior for one spin channel (say spin up), while for the other spin channel (spin down) there is a gap at the Fermi level, so that it behaves either as semiconductor or insulator, depending on the band gap size. When a charge current is passed through such materials spin up electrons move through valence band to conduction band without needing much energy as there is no gap between the valence and conduction bands, while spin down carriers have a gap and therefore it ends up with only one type of spin carriers (spin polarized carriers). Ideally HM materials show 100% spin polarization and therefore are very promising for spintronic applications. Half metallicity of the materials can be checked by mapping the bands by means of angle-resolved photoemission spectroscopy (ARPES) as well as transport measurements (an indirect probe). HM property has been reported in various families of materials such as transition metal chalcogenides (e.g. CrSe) [22], some oxides (e.g. $CrO_2$, $Fe_3O_4$) [23], europium chalcogenides (e.g. EuS), pnictides (e.g. CrAs) [24], double perovskites (e.g. $Sr_2FeReO_6$) [25], manganites (e.g. $La_{0.7}Sr_{0.3}MnO_3$) [23,26], diluted magnetic semiconductors (e.g. Mn doped GaAs) [27] and Heusler alloys (e.g. NiMnSb) [23,28].

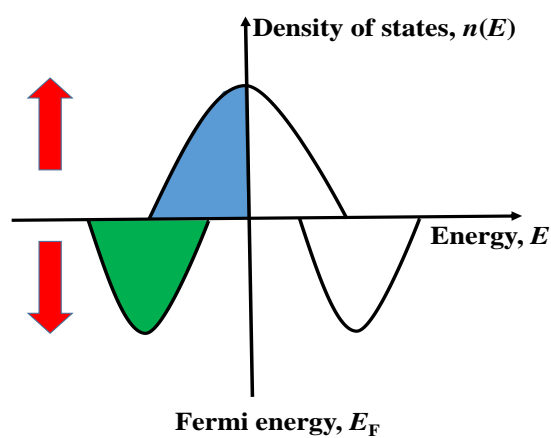

**Figure 3.** The schematic band diagram for half-metallic materials. The arrows represent spin up and down electrons bands. The spin up channel does not have any gap between valence and conduction bands, while spin down has a finite gap.

**3.1.1 Heusler alloys**

Heusler alloys (HAs) are known as multifunctional materials as these materials show diverse magnetic and electronic properties due to their tunable electronic structures and therefore are useful in various applications [29]. HAs are mainly divided in two categories; full Heusler alloys and half-Heusler alloys (also called semi-Heusler alloys). A full HA consists of four interpenetrating *fcc* sublattices and crystallizes in $L2_1$ type crystal structure whereas in half Heusler alloys, one of four *fcc* sublattices is unoccupied, resulting in $C1_b$ type structure. Full HAs have the general formula $X_2YZ$ (XYZ for half Heusler alloys), where X is a transition

S. Gupta, Handbook on the Physics and Chemistry of Rare Earths, Vol. 63, (2023).

metal element, Y is a rare earth or transition metal, and Z is a main group element [30,31]. If the two X in the general formula $X_2YZ$ are different transition metal elements, a quaternary Heusler alloy (with general formula XX'YZ) with a different group of symmetry is formed, which crystallizes in Y-type or LiMgPdSn type structures. Figure 4 represents four interpenetrating *fcc* sublattices and their occupancy for full and half Heusler alloys.

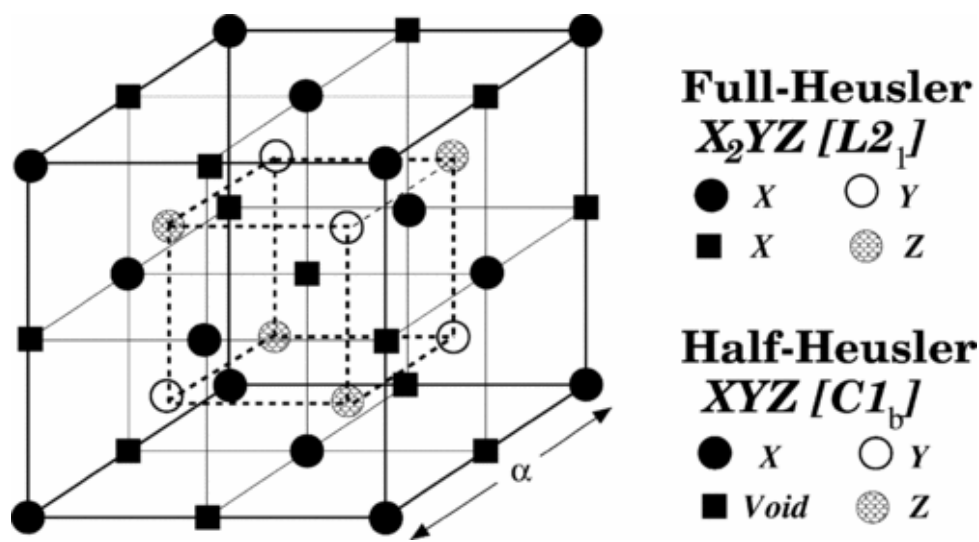

**Figure 4.** Representation of full Heusler alloy ($L2_1$) and Half Heusler ($C1_b$) alloy structures. The lattice consists of four *fcc* sublattices in which one is unoccupied in case of half Heusler alloy. Reproduced with permission from [31] © 2005 American Physical Society.

In 1983, de Groot *et al.* [20] predicted half metallic character of NiMnSb Heusler alloy using first principles electronic structure calculations, which was later confirmed experimentally [32,33]. Though only $CrO_2$ thin films were found to show close to 100% spin polarization experimentally at low temperature, Heusler alloys remain useful for various spintronic applications such as MTJ, GMR devices, spin filters, and spin injection devices [29].

There are several rare earths based Heusler alloys (half, full and quaternary), which have been predicted theoretically and some of them are experimentally realized [34–45]. Some

S. Gupta, Handbook on the Physics and Chemistry of Rare Earths, Vol. 63, (2023).

rare earth HAs are reported to exhibit HM nature [35,36,39,44,45]. Huang *et al.* [35] studied structural, electronic and magnetic properties of rare earth equiatomic quaternary Heusler alloys (EQHA); RXVZ (R = Yb, Lu; X = Fe, Co, Ni; Z = Al, Si) using density functional theory (DFT) calculations.

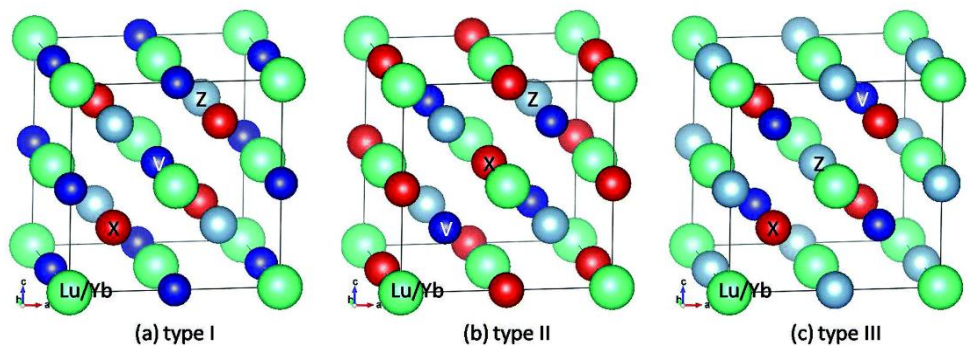

**Figure 5**. Crystal structure of three different types of rare earth equiatomic quaternary Heusler alloys RXVZ (R = Yb, Lu; X = Fe, Co, Ni; Z = Al, Si). Reproduced with permission from [35] © 2021, The Royal Society of Chemistry.

The crystal structure of RXVZ (R = Yb, Lu; X = Fe, Co, Ni; Z = Al, Si) is shown in Figure 5. By fixing rare earth atom position at (0,0,0), there are three different possibilities for occupying the various lattice sites which result in type I, type II and type III crystal structures for RXVZ. The authors [35] theoretically calculated magnetic nature, spin magnetization for individual atoms and in total, and magnetic energies for all the compounds, which are listed in Table 1. Total spin magnetization is compared with the moment expected from the Slater-Pauling rule and the materials showing integer value matching with the one calculated from Slater-Pauling rule are reported to be half-metallic. Furthermore, this can be confirmed from the electronic band structures and DoS plotted in Figure 6. It is clear from DoS plot that the spin up channel shows metallic nature, while the spin down channel shows semiconducting/insulating behavior, confirming half-metallic properties and resulting in 100 % spin polarization in these compounds.

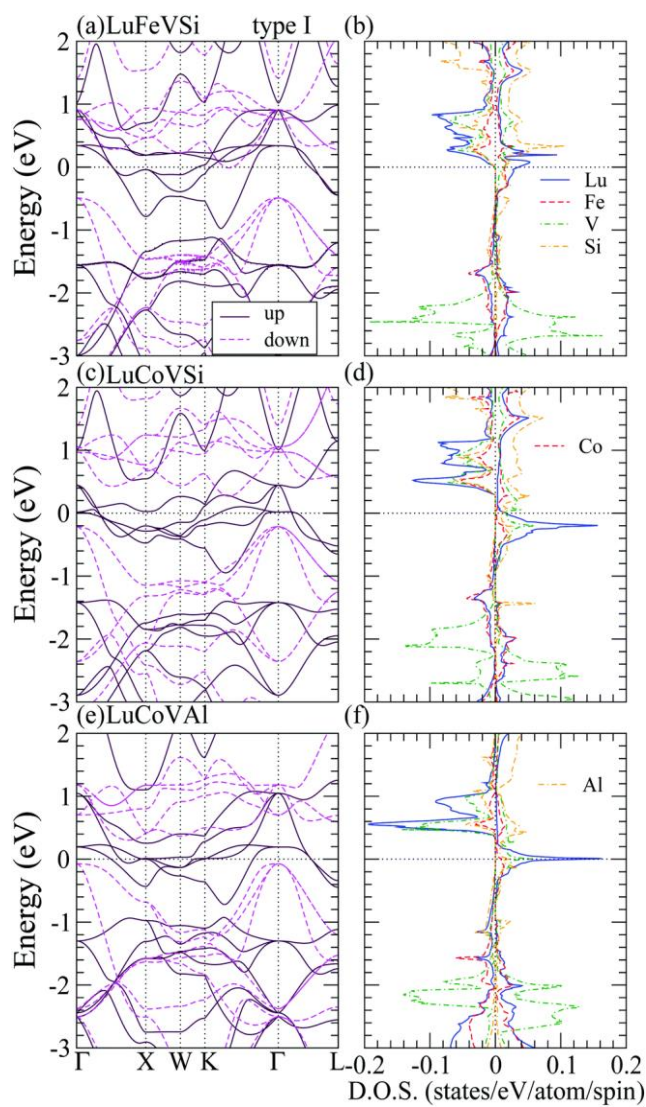

**Figure 6.** The left panels show the electronic band structures, and the right panels show the density of states of equiatomic quaternary Heusler alloys (EQHA) LuFeVSi, LuCoVSi, and LuCoVAl for type I structure. All these EQHA show half metallic behavior. For reference the Fermi level is taken at 0 energy. Reproduced with permission from [35] © 2021, The Royal Society of Chemistry.

S. Gupta, Handbook on the Physics and Chemistry of Rare Earths, Vol. 63, (2023).

**Table 1.** Ground state magnetic nature of rare earth equiatomic quaternary Heusler materials, theoretically calculated total spin moment ($m_t$), spin moment of individual atoms ($m_R$, $m_X$, $m_V$ and $m_Z$ for R, X, V, and Z atoms, respectively), magnetization energy and total magnetic moment ($M_t$) calculated from Slater–Pauling rule. Here FM: ferromagnetic, FiM: ferrimagnetic, HM: half metal, AHM: antiferromagnetic half metal, and MSC: magnetic semiconductor. The table is adapted from Ref. [35].

| RXVZ | Magnetic state | $m_t$ ($\mu_B$) | $M_R$ ($\mu_B$/atom) | $M_X$ ($\mu_B$/atom) | $m_V$ ($\mu_B$/atom) | $m_Z$ ($\mu_B$/atom) | EM (eV/f.u.) | $M_t$ ($\mu_B$) |
|---|---|---|---|---|---|---|---|---|
| YbFeVAl | FiM | 0.06 | -0.11 | 2.16 | -1.85 | 0.023 | -0.141 | 0 |
| YbCoVAl | HM | 1.00 | 0.06 | -0.87 | 1.68 | -0.066 | -0.194 | 1 |
| YbNiVAl | FM | 2.13 | 0.07 | -0.09 | 2.09 | -0.115 | -0.518 | 2 |
| YbFeVSi | HM | 1.00 | 0.13 | -1.49 | 2.03 | -0.008 | -0.215 | 1 |
| YbCoVSi | HM | 2.00 | 0.12 | -0.51 | 2.12 | -0.046 | -0.340 | 2 |
| YbNiVSi | HM | 3.00 | 0.15 | 0.16 | 2.42 | -0.070 | -0.773 | 3 |
| LuFeVAl | NM | 0.00 | | | | | | 1 |
| LuCoVAl | HM | 2.00 | 0.30 | -0.40 | 1.87 | -0.037 | -0.310 | 2 |
| LuNiVAl | HM | 3.04 | 0.34 | 0.25 | 2.23 | -0.058 | -0.698 | 3 |
| LuFeVSi | HM | 2.00 | 0.38 | -0.48 | 1.80 | 0.003 | -0.116 | 2 |
| LuCoVSi | HM | 3.00 | 0.37 | 0.50 | 1.93 | -0.025 | -0.399 | 3 |
| LuNiVSi | FM | 3.99 | 0.50 | 0.57 | 2.47 | 0.004 | -0.720 | 4 |
| YbFeVAl | AHM | 0.00 | -0.04 | 2.56 | -2.07 | -0.012 | -0.565 | 0 |
| YbCoVAl | FM | 2.82 | -0.03 | 0.99 | 1.64 | -0.001 | -0.171 | 1 |
| YbNiVAl | FM | 2.91 | 0.02 | 0.32 | 2.15 | 0.043 | -0.440 | 2 |
| YbFeVSi | FiM | 0.58 | -0.02 | 2.62 | -1.87 | 0.133 | -0.703 | 1 |
| YbCoVSi | FM | 2.99 | -0.02 | 1.19 | 1.59 | 0.013 | -0.313 | 2 |
| YbNiVSi | FM | 2.27 | 0.01 | 0.19 | 1.87 | -0.049 | -0.431 | 3 |
| LuFeVAl | FM | 1.43 | -0.07 | 0.33 | 1.11 | -0.018 | -0.056 | 1 |
| LuCoVAl | FM | 2.98 | -0.07 | 1.22 | 1.64 | 0.026 | -0.366 | 2 |
| LuNiVAl | FM | 2.50 | -0.04 | 0.33 | 1.97 | 0.003 | -0.439 | 3 |
| LuFeVSi | FM | 1.82 | 0.01 | 2.34 | -0.53 | 0.067 | -0.455 | 2 |
| LuCoVSi | FM | 2.96 | -0.03 | 1.26 | 1.53 | 0.030 | -0.349 | 3 |
| LuNiVSi | FM | 2.43 | 0.05 | 0.19 | 1.93 | -0.056 | -0.294 | 4 |
| YbFeVAl | FM | 1.72 | -0.06 | 2.06 | -0.17 | -0.037 | -0.131 | 0 |
| YbCoVAl | FM | 1.80 | 0.10 | -0.53 | 2.05 | -0.015 | -0.417 | 1 |
| YbNiVAl | FM | 2.33 | 0.08 | -0.10 | 2.20 | -0.052 | -0.651 | 2 |
| YbFeVSi | HM | 1.00 | 0.24 | -1.82 | 2.29 | -0.009 | -0.400 | 1 |

| YbCoVSi | FM  | 2.12 | 0.12 | -0.16 | 2.02 | -0.047 | -0.415 | 2 |
| YbNiVSi | FM  | 3.00 | 0.26 | -0.03 | 2.48 | -0.042 | -0.559 | 3 |
| LuFeVAl | FM  | 2.83 | 0.11 | 0.77  | 1.78 | -0.031 | -0.230 | 1 |
| LuCoVAl | FM  | 2.20 | 0.15 | -0.24 | 2.10 | -0.025 | -0.581 | 2 |
| LuNiVAl | FM  | 3.00 | 0.23 | -0.01 | 2.42 | -0.013 | -0.834 | 3 |
| LuFeVSi | HM  | 2.00 | 0.27 | -0.73 | 2.15 | -0.002 | -0.412 | 2 |
| LuCoVSi | MSC | 3.00 | 0.26 | 0.06  | 2.34 | -0.011 | -0.656 | 3 |
| LuNiVSi | FM  | 3.70 | 0.35 | 0.32  | 2.56 | 0.002  | -0.432 | 4 |

### 3.2 Topological Insulators

In 1982, Thouless *et. al.* introduced the concept of topological order [46]. More than 30 years later, in 2005, Kane and Mele [47] realized that spin-orbit interaction can lead to a new topological insulating electronic phase, which has emerged as a new field in condensed matter physics. Since then, many researchers have theoretically predicted topological insulating phases and observed them experimentally in various materials [48–54]. Topological insulators (TIs) are quantum materials having insulating bulk and conducting surface (edge in two dimensions) states, which are protected by time-reversal symmetry (TRS) [55]. The TRS protected non-trivial topological surface is characterized by two linearly dispersing states having correspondence between bulk valence and conduction bands. The point where these linearly dispersing states cross each other is called Dirac point. Carriers on topologically non-trivial surface states are massless fermions. Due to strong SOC, surface carriers show property of spin-momentum locking, meaning that the direction of spin is univocally determined by the direction towards which carriers are traveling [56]. Gapless topological surface (in 3D) and edge (in 2D) states are obtained through band inversion (Figure 7) derived by relativistic spin-orbit interactions, which naturally suppresses backscattering of carriers. In principle, this could lead to 100% spin polarization and high carrier mobility. Therefore, in addition to their fundamental interests, TIs are expected to have applications ranging from thermoelectric, to spintronics and quantum computation [46,56–64].

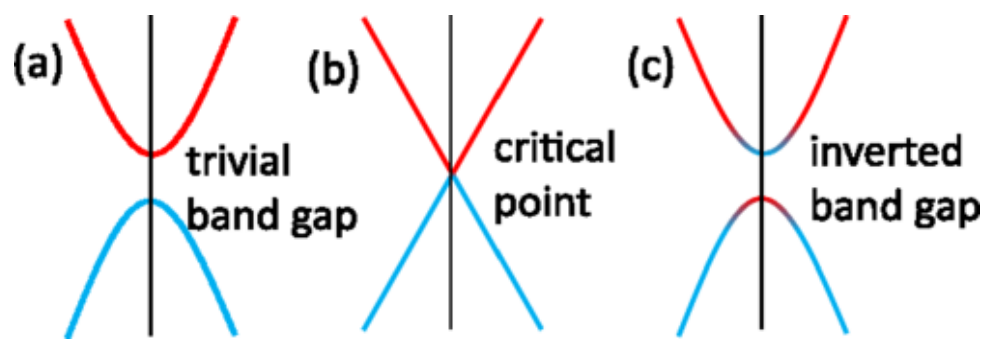

**Figure 7.** Schematic band diagrams: (a) band gap in trivial insulator, (b) the band gap closes at a critical point, (c) bands get inverted and reopen, resulting in non-trivial topological state. Reproduced with permission from [55] © 2016 American Physical Society.

The main ingredients for a material to be a topological insulator are (i) bulk band gap, (ii) appropriate magnitude of spin orbit interactions (to invert bands), (iii) time reversal symmetry. However, in some cases, it has been observed that the introduction of strain can also tune the electronic band structures of a material and result in band inversion [48]. As we know, Heusler alloys show tunable electronic band structures. Selecting materials with variable lattice parameters (growth of material on suitable substrate or by applying pressure) and with large SOC strength (by substitution of elements), allows one to tune the bandgap and invert the band structures [42]. RE-based half Heusler alloys are good candidates to be explored as they fulfil both requirements.

Non-trivial TSS were reported experimentally in some of these materials using ARPES technique and other indirect measurements such as Shubnikov-de Haas (SdH) oscillations and nuclear magnetic resonance studies [37,41].

S. Gupta, Handbook on the Physics and Chemistry of Rare Earths, Vol. 63, (2023).

**3.2.1 Angle-resolved photoemission spectroscopy (ARPES)**

ARPES is a technique which can directly probe the allowed energy and momenta along with the spin of an electron in a given material. By mapping these properties, the technique can provide detailed information about the electronic band structure and Fermi surface of this material.

Liu *et al.* used single crystals of LnPtBi (Ln = Lu, Y) with either (111) or (001) surfaces and performed ARPES measurements [41]. To understand experimental results in detail, the authors also performed band structure calculations using two different methods (slab model and recursive Green's function) for Bi-terminated LuPtBi (111) surface, which show X shape TSS (Figure 8 a,b) and are in agreement with the experimental results (Figure 8 c-f) [41]. By using these two methods, the authors could separate out trivial surface states resulting from the dangling bonds at the surface from non-trivial TSS resulting in X shape band dispersions as shown in Figure 8(b). The other material YPtBi also displays similar TSS for which both calculations and experimental observations agree excellently well (for details, please see Ref. [41]).

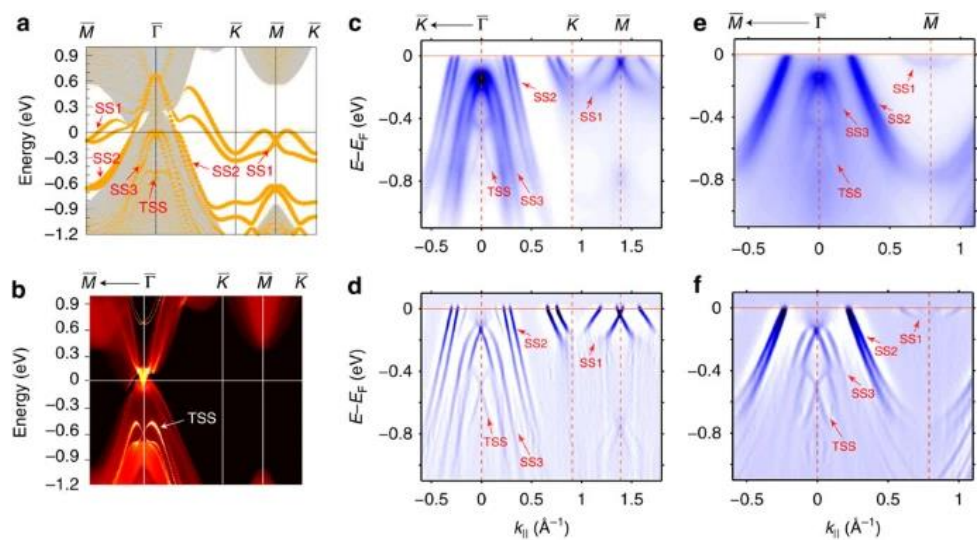

**Figure 8.** Observation of topological surface state (TSS) in LuPtBi (111): calculated electronic band structures for LuPtBi (111) Bi terminated surface with (a) both metallic and non-trivial

S. Gupta, Handbook on the Physics and Chemistry of Rare Earths, Vol. 63, (2023).

topological surface states are shown (b) only non-trivial topological surface states. (c) plot of photoemission intensity, (d) the second derivative plot of photoemission intensity along $\Gamma - K - M$ direction. (e) plot of photoemission intensity and (f) the second derivative plot along $\Gamma - M$ direction. SS stands for topological trivial metallic surface originated from dangling bonds on the surface of the sample. Reproduced with permission from [41] © 2016 Springer Nature.

**3.2.2 Shubnikov-de Haas (SdH) oscillations**

The electronic band structures of topological materials can be directly measured by ARPES; however, for these measurements, materials should be cleaved in the desired direction, which sometimes is difficult. Also, ARPES is a sophisticated technique and not a common laboratory technique. The indirect ways to establish the topological behavior of materials is to perform transport measurements and map Fermi surface determination by measuring Shubnikov-de Haas (SdH) oscillations. For more details about SdH, one can refer to Ref. [65]. Various RE half Heusler alloys such LnPtBi (Ln= Y, Tb, Gd, Er, Lu) [66–69] and YPdBi [70] were found to show SdH oscillations. The temperature dependence of electrical resistivity of these materials (except LuPtBi) shows semi metallic behavior down to ~50 K. As an example, the SdH oscillations in YPtBi are depicted in Figure 9. The parameters determined from the fitting of SdH oscillations data are given in Table 2 [69].

S. Gupta, Handbook on the Physics and Chemistry of Rare Earths, Vol. 63, (2023).

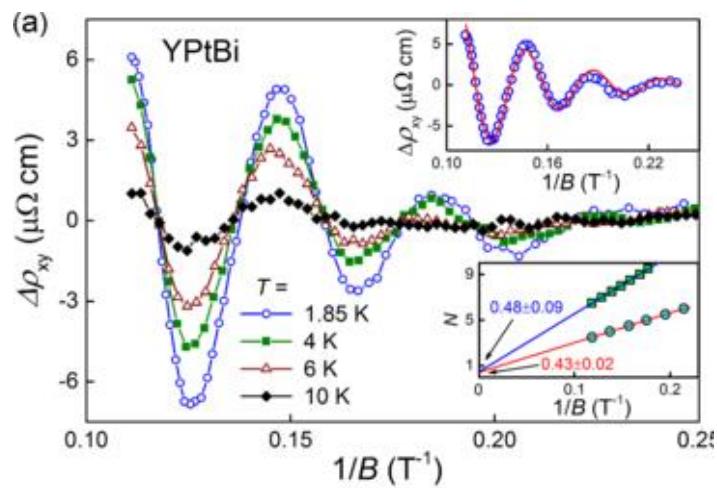

**Figure 9.** Shubnikov–de Haas (SdH) oscillations of $\rho_{xy}$ as a function of temperature. The upper inset shows Lifshitz-Kosevich (LK) fit to SdH oscillations recorded at $T$ = 1.85 K and the lower inset shows Landau Level index plots fitted with straight lines (circles for $\rho_{xy}$ and square for $\rho_{xx}$). Reproduced with permission from [69] © 2016, American Physical Society.

**Table 2.** Value of various parameters determined from the analysis of SdH oscillations for YPtBi. The table is adapted from Ref. [69].

| Parameters | Value |
|---|---|
| Berry phase | 0.75 $\pi$ |
| Mean free path | 26.8 nm |
| Surface scattering time | 1.88 $\times 10^{-13}$ s |
| Fermi vector | 2.74$\times 10^6$ /cm |
| Cyclotron mass | 0.22 $m_e$ |

S. Gupta, Handbook on the Physics and Chemistry of Rare Earths, Vol. 63, (2023).

From Table 2, it can be noted that the effective electron mass reflects the linear dispersion of the electronic band structure. The $0.75\pi$ value of Berry phase is non-trivial and is close to the one expected for Dirac fermions.

When topological properties are combined with magnetism, this can lead to new quantum states such as quantum anomalous Hall effect (QAHE), which was first discovered in Cr-doped $(Bi,Sb)_2Te_3$ [71]. Magnetism can be introduced into topological insulators in different ways; (i) by doping magnetic impurities into intrinsic TIs, (ii) by proximity effect and (iii) by synthesizing magnetic TIs [72]. By introducing magnetism into TI, time-reversal symmetry is broken, which induces a bandgap in the topological surface states (Figure 10) leading to 1-dimensional chiral edge conduction [72]. The gap size in TSS depends on the magnitude of magnetic moment.

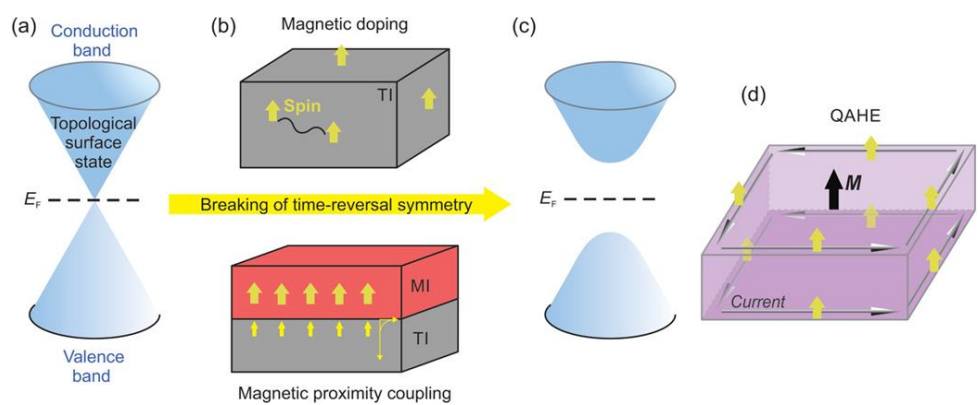

**Figure 10.** Scheme showing the effect of magnetism introduction in a topological insulator. a) Topological surface states with Fermi energy ($E_F$) at Dirac point b) Time reversal symmetry breaking upon introduction of magnetism. c) opening of a gap at Dirac point after magnetism introduction. d) Observation of quantum anomalous Hall effect with representation of spin momentum locking. Reproduced under the terms Creative Commons CC BY license (https://creativecommons.org/licenses/) from [72] © 2021 Wiley-VCH Verlag GmbH.

S. Gupta, Handbook on the Physics and Chemistry of Rare Earths, Vol. 63, (2023).

It was theoretically predicted that doping of magnetic impurities such as magnetic transition metals can induce QAHE in TIs [73]. Following this, a systematic study on structural, magnetic and electronic properties have been carried out to explore magnetic-doped TIs grown by molecular beam epitaxy (MBE) [63,71,74,75]. Cr-doped $Sb_2Te_3$ was reported to be the highest Curie temperature (~250 K) magnetic-doped TI [63]. The main challenge in magnetic topological insulators (MTIs) was to increase the QAHE temperature. As transition metals have small atomic moments, increasing this temperature requires larger TM doping, which further introduces surface disorder. Therefore, rare earths having large atomic moments are employed as magnetic dopants into TI to raise the QAHE temperature [76]. Doping of various REs in TIs has been investigated and it was observed that though there is no observation of QAHE in these systems, they nevertheless display interesting properties. Table 3 lists RE-doped TIs, which show paramagnetic (PM), antiferromagnetic (AFM), and ferromagnetic (FM) nature, when doped into intrinsic TIs. Paramagnetic to antiferromagnetic transition was observed in Gd doped MTI, the possible reason for the magnetic order being Gd-Gd coupling through chalcogen ions [77–79]. Topological insulators doped with Sm have ferromagnetic nature, and are also expected to be magnetic axion insulator systems for inducing surface QAHE and chiral hinge states [72,80].

**Table 3.** Various rare earth doped topological insulators, their doping concentration, magnetic transition temperature, magnetic nature, observed effective moment per rare earth ion.

| Rare earth doped TI | Doping concentration | Magnetic transition temperature | Magnetic nature | Effective magnetic moment ($\mu_{eff}$) | Ref. |
|---|---|---|---|---|---|
| $Bi_{1.09}Gd_{0.06}Sb_{0.85}Te_3$ | - | 4-8 K | AFM | 8.1 $\mu_B$ | [79] |
| $Bi_{2-x}Gd_xTe_3$ | $0 \leq x \leq 0.2$ | 12 K for x= 0.2 | AFM | 7.06 $\mu_B$ for x= 0.2 | [78] |
| $(Ho_xBi_{1-x})_2Te_3$ | 0.14 | - | AFM | ~10.6 $\mu_B$ | [81] |
| $(Gd_xBi_{2-x})Se_3$ | 0.3 | 6 | AFM | 2.46 $\mu_B$ | [77] |
| $(Sm_xBi_{1-x})_2Se_3$ | 0.05 | 52 | FM | - | [80] |
| $(Dy_xBi_{1-x})_2Te_3$ | $x = 0- 0.355$ | - | PM | 12.6 $\mu_B$ for x= 0.023  ~4.3 $\mu_B$ for x= 0.355 | [82] |

Doping of magnetic impurities into a TI not only introduces magnetism to it but also creates crystal defects/disorder, impurity states into the insulating gap, as well as magnetic scattering centers, which deteriorate transport properties of surface carriers and their mobility [83]. To make the TI surface free from impurities, magnetic proximity effect has been considered an alternative approach to induce magnetism in TIs and break time-reversal symmetry [72,84,85]. Moreover, this approach is also advantageous to confine Majorana fermions in topological superconductors as the proximity coupling of FM insulators lifts spin degeneracy without disturbing Cooper pairing [86]. Europium sulfide, EuS is a rare-earth based ferromagnetic insulator with Curie temperature of ~17 K; it has been used for magnetic proximity effect in $Bi_2Se_3$ TI thin films by forming a $Bi_2Se_3$/EuS heterostructure [83]. The magnetic and transport properties of bi-layered $Bi_2Se_3$ [0001]/EuS[111] is displayed in Figure 11.

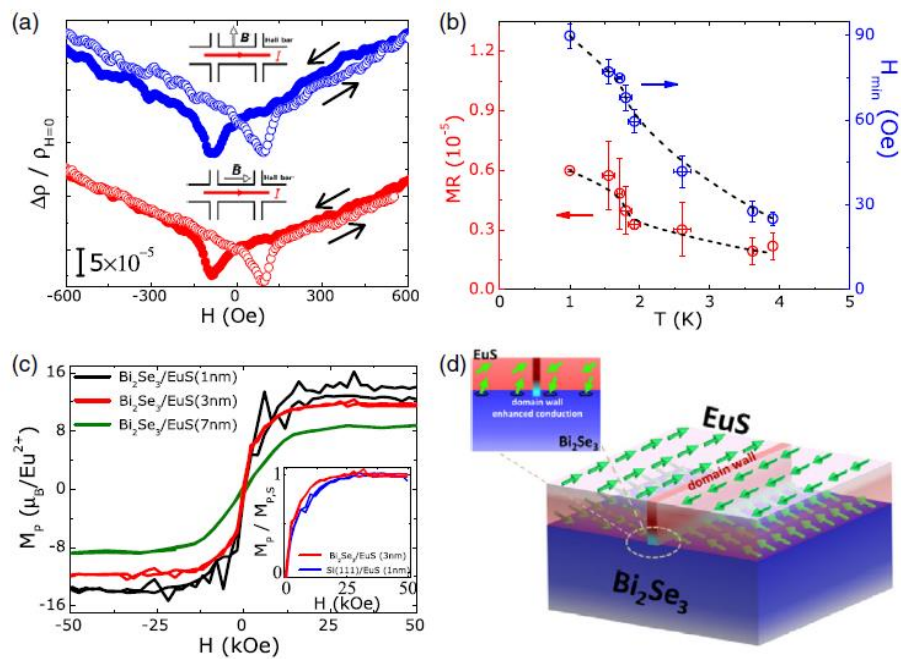

**Figure 11**. (a) In-plane magnetic field ($H$) dependence of planar magneto-resistivity of $Bi_2Se_3$/EuS(1 nm) at $T$= 1 K. The inset shows schematic of Hall bar with current and magnetic field directions. (b) Temperature dependence of magnitude of magneto-resistivity dip and coercive field ($H_c$). (c) $H$ dependence of perpendicular magnetic moment ($M_P$) per $Eu^{2+}$ ion. (d) Scheme showing magnetic moment distribution at the $Bi_2Se_3$/EuS interface. Reproduced with permission from [83] © 2013 American Physical Society.

As it can be clearly seen from Figure 11(c), $Bi_2Se_3$ gets magnetized by EuS magnetic proximity coupling, which breaks time-reversal symmetry in the topological insulator. The signature of FM order is also seen in electrical transport. The system shows sharp hysteretic dips in resistance at 1K, which appear once in the direction of field sweep (Fig. 11 a). This study demonstrate that $Bi_2Se_3$/EuS exhibits proximity-induced interfacial magnetism, which arises due to the uniform exchange field over topological surface states without disturbing the TI structure [83]. Apart from EuS, magnetic proximity coupling effect in TIs has also been studied in various other RE-based magnetic insulators such as $Tm_3Fe_5O_{12}$ (TIG) [87,88] and $Y_3Fe_5O_{12}$ (YIG) [88–92]. Proximity effects in their heterostructures can develop various technological advanced devices. For example, by fabricating heterostructures with TI sandwiched with FM insulator for zero-field quantum Hall edge channels and heterostructures holding Majorana bound states by using FM insulator at TI superconductor interfaces [83,86].

Though the magnetic proximity effect can induce magnetism in topological insulators, it is effective up to few nm down the interface only and therefore it cannot guarantee the involvement of topological surface state carriers in mediating the coupling between local moments [72]. Magnetic materials with TSS are better candidates from this point of view as there is no need of doping and proximity effect. It has been reported that several RE-based Heusler alloys are predicted to have an electronic band structure which is borderline with that of topological insulators. Electronic structures of these materials can be easily tuned from trivial to TI by varying lattice parameters (applying pressure or using suitable substrate) or by strength of spin-orbit coupling [42]. Furthermore, devices built from these materials provide more ways to tune electronic structures from trivial materials to topological insulators by electrical gating or fabricating quantum well structures. RE Heusler alloys show a variety of phenomena such as heavy fermion behavior, magnetism and superconductivity, for instance, and therefore combining these phenomena with non-trivial topological states can lead to various new topological effects such as axions, image monopole effect and topological superconductivity [42]. For example, the bulk magnetism in LnPtBi (Ln= Nd, Sm, Gd, Tb, Dy) could lead to realization of dynamical axion, a spin-wave exciton coupled topologically with electromagnetic field [42,93]. YPtBi displays heavy fermion behavior, which could lead to realization of topological Kondo insulators [94] while superconducting properties of LaPtBi, which is a low carrier non-centrosymmetric system, are predicted to support topological superconductivity [95].

Chadov *et al.* performed first principles electronic band structure calculations and reported that around fifty Heusler alloys including many RE-based Heusler alloys exhibit inverted band structures like the two-dimensional topological insulator HgTe [42,95]. The authors calculated the energy of the $\Gamma_6$ and $\Gamma_8$ bands and plotted the difference energy between both bands as a function of lattice constant and strength of spin-orbit interactions for the Heusler alloys containing Sc, Y, La, and Lu, along with HgTe (a TI) and CdTe for comparison as shown in Figure 12 [42]. It can be noted that compounds with $E_{\Gamma_6} - E_{\Gamma_8} > 0$ are trivial insulators whereas those with $E_{\Gamma_6} - E_{\Gamma_8} < 0$ are non-trivial topological insulator candidates. It can be seen from Figure 12 that LnPtBi and LnAuPb series members always exhibit inverted band structures, a property required for topological insulators.

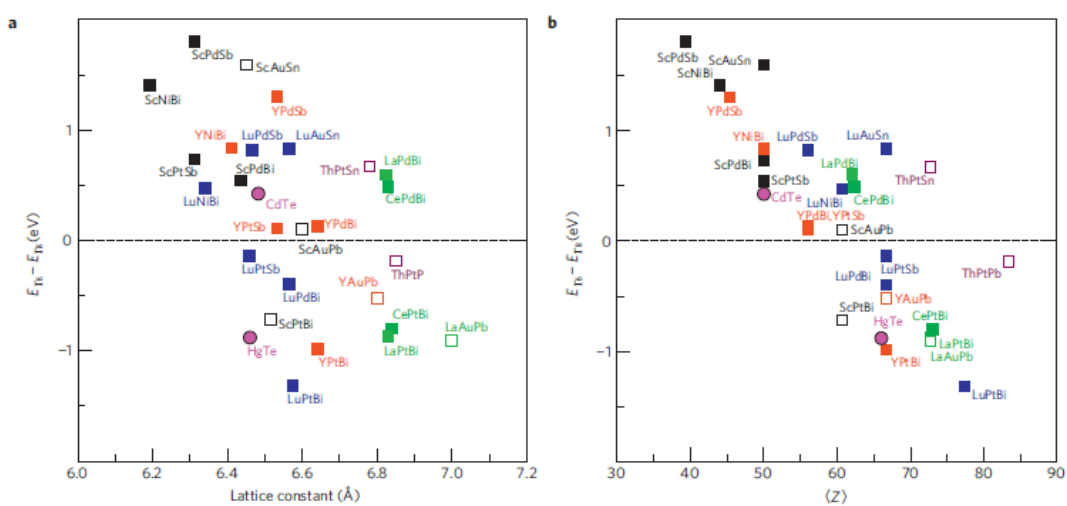

**Figure 12.** $E_{\Gamma_6} - E_{\Gamma_8}$ calculated as a function of experimental lattice constants and spin orbit interaction strength for various Heusler alloys. (a) $E_{\Gamma_6} - E_{\Gamma_8}$ as a function of experimental lattice constants. (b) $E_{\Gamma_6} - E_{\Gamma_8}$ as a function of average spin orbit coupling strength, $<Z> = \left(\frac{1}{(N)}\right)\sum_{i=1}^{N} Z(X_i)$, where $N$ is 2 for binary and 3 for ternary. $E_{\Gamma_6} - E_{\Gamma_8}$ for HgTe and CdTe are included for reference. Reproduced with permission from [42] © 2010 Springer Nature.

### 3.3 Dirac and Weyl semi-metallic materials

There are various non-trivial topological materials in addition to the topological insulators. Dirac and Weyl semimetals are also examples of non-trivial topological matters, however instead of insulators, these materials are topological semimetals [96–98]. In topological Dirac semimetals (TDS), the bulk valence and conduction bands meet at Dirac point and disperse linearly in all momentum directions, which resembles as a natural 3D counterpart of graphene [96]. Topological Dirac semimetals are similar to the topological Weyl semimetal (TWS); however, TDS are equivalent to two merged compensated Weyl points (i.e. the crossing involves four bands rather than two bands) [99]. Due to their unique electronic structures, TDS show many unusual physical phenomena for example; quantized magnetoresistance, giant diamagnetism, and oscillating quantum spin Hall effect in quantum well structures [96]. Topological Weyl semimetals represent another, novel quantum state, which exhibits a pressure-induced anomalous Hall effect and its quantum version when confined in a quantum well structure.

A number of rare earth materials such as GdPtBi [68], PrAlGe [100], SmAlSi [101], and REFeAsO [102] have been described experimentally as TWSs. GdPtBi is a half Heusler alloy, characterized by a zero-gap semiconducting nature with quadratic electronic band structure. When a magnetic field is applied, the Zeeman energy induces Weyl nodes [68]. It shows large negative longitudinal magnetoresistance with field steering properties suggesting the presence of chiral anomaly. It has also been observed that the chiral anomaly strongly suppresses thermopower in this material. Figure 13 shows temperature and field dependence of transport measurements for a GdPtBi crystal cleaved along the [110] direction. Negative field dependence of longitudinal resistance below 150 K is clearly observed in Figure 13(b), suggesting that chiral anomaly could be the possible reason for the large negative longitudinal magnetoresistance.

S. Gupta, Handbook on the Physics and Chemistry of Rare Earths, Vol. 63, (2023).

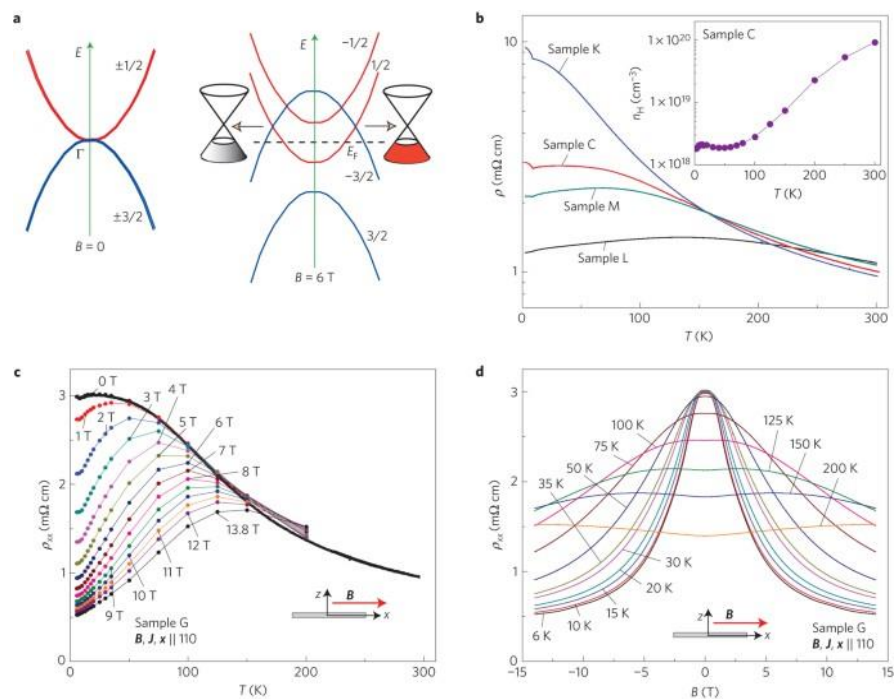

**Figure 13**. (a) Schematics of bands with and without magnetic field and their splitting. (b) Zero-field electrical resistivity for different samples of GdPtBi as a function of temperature. The inset displays the Hall carrier density in sample C. (c) Temperature dependence of longitudinal resistivity $\rho_{xx}$ [when the magnetic field is parallel to current and sample direction (110)] (d) Field dependence of $\rho_{xx}$, showing a bell-shaped profile, which strongly supports the chiral anomaly as the origin of the longitudinal magnetoresistance. Reproduced with permission from [68] © 2016, Springer Nature.

Chamorro *et al*. [103] reported Dirac fermion behavior in the layered Sb square-lattice material LaCuSb$_2$. The ARPES measurements confirm linearly dispersing band crossing electronic band structures, which is necessary to generate Dirac fermions. The transport measurements carried out on the sample evidence weak antilocalization as well as linear field dependence, indicative of a non-trivial topological phase. Moreover, observation of Shubnikov-de Hass quantum oscillations and the determined effective electron mass of 0.065m$_e$ support the idea that the material is really a topological Dirac semimetal [103].

**3.4 Fully compensated ferrimagnets**

Ferrimagnets consist of inequivalent magnetic sublattices with antiferromagnetic coupling between them and therefore can benefit from the properties of both ferromagnets and antiferromagnets [104]. Ferrimagnets have been employed traditionally in various applications such as microwave [105], spin-wave devices [106] and magneto-optical recording [107]. These applications essentially rely on the low moment of ferrimagnets, ignoring the antiferromagnet like properties arising from antiferromagnetic coupling between their sublattices [104]. Recently fully compensated ferrimagnet (FCF) research has drawn significant attention. Antiferromagnetic coupling of magnetic sublattices in carefully selected ferrimagnets cancels moments from different sublattices, resulting in vanishing magnetic moments [104,108–110]. These materials with vanishing moments and antiferromagnetic coupling have many advantages [111,112]; (i) AFM (antiparallel) coupling in FCFs make them operate at a much higher frequency (up to terahertz) compared to that of ferromagnets (gigahertz), enabling them to perform ultrafast data operation; (ii) vanishing moments in these materials result in negligible stray field, making them more robust against magnetic perturbation fields and therefore they are very promising for next generation data storage devices; (iii) as critical current for magnetization switching is proportional to the strength saturation magnetization, due to smaller critical currents, these materials are attractive for spin transfer torque spintronic devices; (iv) unlike antiferromagnets, FCFs show finite Zeeman coupling and spin polarization. Therefore, magnetization in FCFs can be easily controlled by external magnetic field. Furthermore, spin polarization in these materials opens up many directions for spin based research [104]. AFM coupling of non-equivalent magnetic sublattices in ferrimagnets offer great tunability of the gyromagnetic ratio ($\gamma$), which is described as the ratio of magnetization ($M$) to angular momentum density ($A$). Both $M$ and $A$ can be almost independently tuned by the variation of temperature and /or materials composition. The tuning of $M$ and $A$ can pass through a trajectory with two remarkable points; one corresponds to an angular momentum

S. Gupta, Handbook on the Physics and Chemistry of Rare Earths, Vol. 63, (2023).

compensation point ($T_A$), where the total angular momentum vanishes and the other is a magnetization compensation point ($T_M$), where the total magnetization vanishes. At these points, ferrimagnets have been reported to show many interesting behaviors as can be seen in various experiments [104,113,114].

RE-TM ferrimagnets, for example GdFeCo, TbFeCo, TbCo, GdCo, (Sm,Nd)ScGe, feature remarkable spintronic properties [104]. In RE, $5d$ electrons behave as conduction electrons and play a very important role by mediating the exchange interactions [2]. These $5d$ spins in RE interact antiferromagnetically with the $3d$ spins in TM at the Fermi level [2]. This mechanism is responsible for unique electronic, optical, and spintronic properties of RE-TM ferrimagnets. Experimental observation of all optical magnetization switching in GdFeCo was one of novel discoveries in ferrimagnets. The authors [113] carried out X-ray magnetic circular dichroism measurements to study magnetization switching in GdFeCo. After optical excitation, the magnetization of both the sublattices Gd and Fe decreases rapidly but in different manner: Fe magnetization falls off within 0.2 picosecond, while for Gd it takes 1.5 picoseconds; in addition, their magnetization flips directions and is regained at significantly different timescales. As a consequence, for a very short time, Gd and Fe moments show parallel alignment despite having AFM coupling in their ground state, as shown in Figure 14 [113].

S. Gupta, Handbook on the Physics and Chemistry of Rare Earths, Vol. 63, (2023).

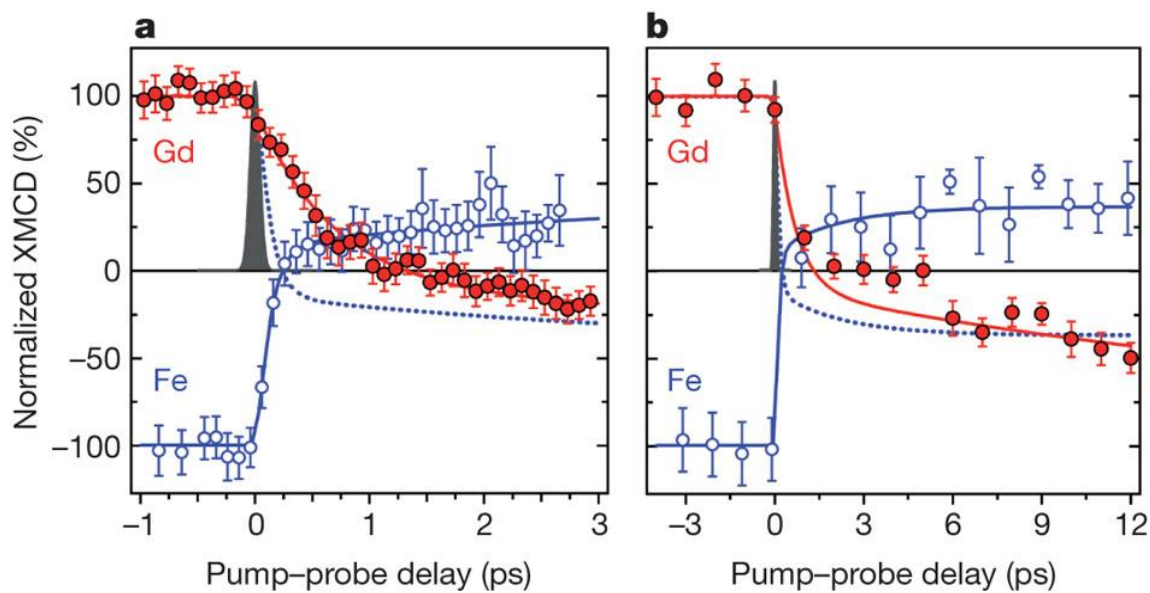

**Figure 14.** Time resolved X-ray magnetic circular dichroism pumped with a linearly polarized femtosecond laser pulse in GdFeCo, showing transient dynamics of Gd and Fe magnetic moments within (a) the first 3 ps (b) and 12 ps timescale. Reproduced with permission from [113] © 2011, Springer Nature.

It has been predicted that AFM spin dynamics emerge much faster than that of the FM counterpart [115]. However, due to lesser magnetic field sensitivity of antiferromagnets, experimental investigations of spin dynamics in these materials have remained scarce. Recently Kim *et al.* [114] experimentally realized fast field-driven AFM spin dynamics in GdFeCo at $T_A$. The remarkable enhancement of field-driven domain wall (DW) mobility up to 20 km/sT (see Figure 15) is a consequence of AFM spin dynamics at $T_A$ supported by collective coordinate approach generalized for ferrimagnets and atomistic spin model simulations [114]. A spin orbit torque (SOT) assisted DW movement near $T_A$ (~260 K) has been reported in another RE-TM ferrimagnet, GdCo [116]. One advantage of SOT driven DW motion over field-driven is that SOT can drive the DW at all the temperatures even at $T_M$, where field driven DW motion is not feasible [104].

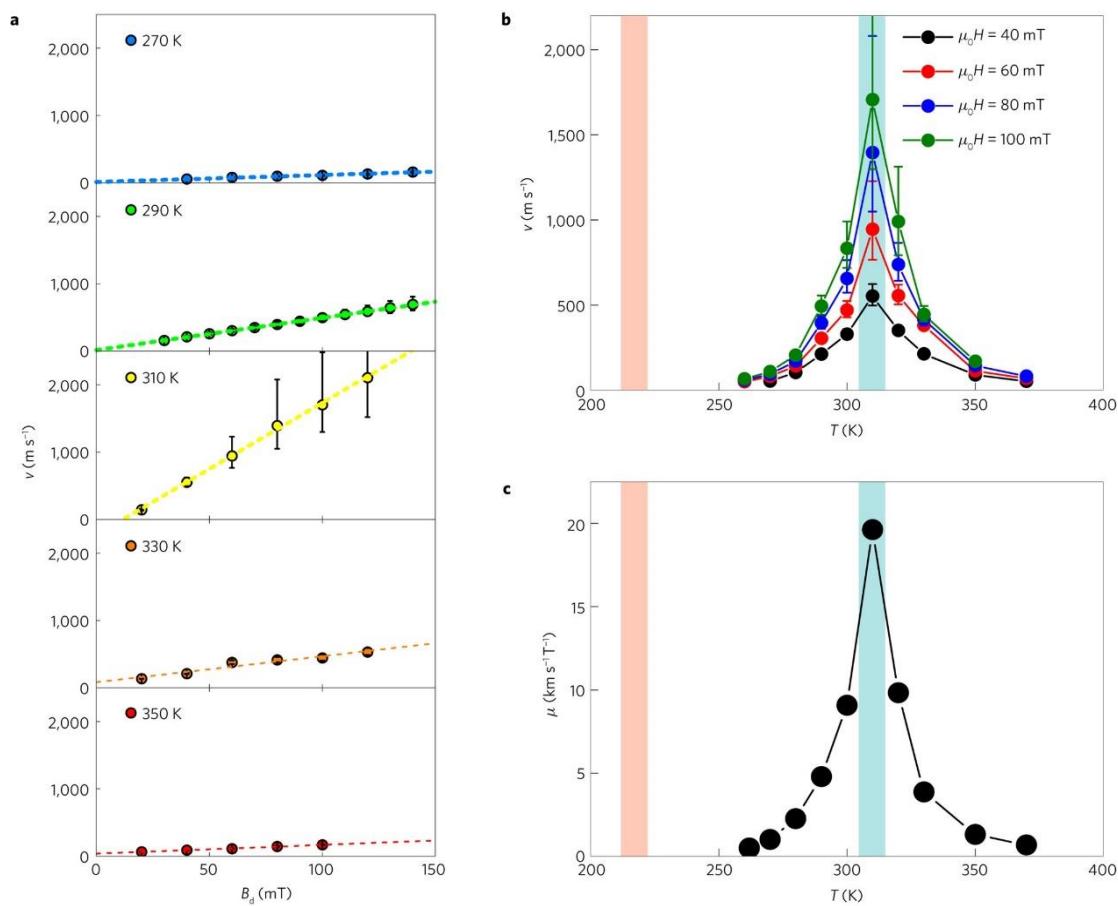

**Figure 15.** (a) Driving field ($B_d$) dependence of domain wall speed ($v$) as a function of temperature. Dashed lines in the figure show best fit of function $v = \mu(B_d - B_0)$. (b) Temperature dependence of $v$ as a function of $B_d$. (c) Temperature dependence of domain wall mobility ($\mu$). The blue and red shaded region in Fig. (b) and (c) represent the angular momentum compensation and magnetization compensation temperatures. Reproduced with permission from [114] © 2017, Springer Nature.

Fully compensated ferrimagnets also feature interesting spin transport properties due to non-zero spin polarization at both $T_A$ and $T_M$ [104,117]. It has been suggested theoretically that in antiferromagnets, the spin coherence length, $\lambda_c$ could be longer than the ferromagnets due to staggered spin order on the atomic scale [118–120]. In order to confirm the theoretical prediction of long $\lambda_c$, Yu *et al.* [121] recently performed an experiment with Co/Tb ferrimagnetic system. The authors chose a Co/Tb system instead of antiferromagnets for two reasons; (i) Co/Tb shows AFM coupling between Co and Tb, can have longer $\lambda_c$ and exhibits a bulk-like torque characteristics unlike FMs which exhibit a surface torque [121]. A bulk-like torque in antiferromagnets can be explained semi-classically: an electron spin in an antiferromagnet has exchange interaction with an alternating spin orientation on atomic scale, which can be averaged to zero over two sublattices and thus is expected to have infinitely long $\lambda_c$, resulting in bulk like torque characteristic [121]. (ii) The second reason the authors chose ferrimagnet is that it has non-zero magnetic moment and therefore models established for a ferromagnet can be extended to a ferrimagnet [114]. The authors performed spin pumping experiments on two multilayer systems; [Co/Tb] and [Co/Ni] with a Co capping layer as shown in Figure 16(a) and determined $\lambda_c$. The spin current is generated by Co ferromagnet, which enters into the [Co/Tb] or [Co/Ni] multilayer system through a Cu layer and is detected at the bottom Pt layer via inverse spin Hall voltage ($V_{ISHE}$). The latter can be measured only if $\lambda_c$ is longer than the thickness of the multilayer structures. As shown in Figure 16(c) significant spin-pumping induced $V_{ISHE}$ is detected for the [Co/Tb] multilayer system whereas almost no $V_{ISHE}$ is seen in the [Co/Ni] multilayer system, which is ferromagnetic in nature [121]. This confirms that $\lambda_c$ is longer in the [Co/Tb] ferrimagnetic system compared to the [Co/Ni] ferromagnetic system. Another experiment performed on a CoGd ferrimagnetic alloy also evidenced a long $\lambda_c$, which is four times larger than reported in ferromagnets [122].

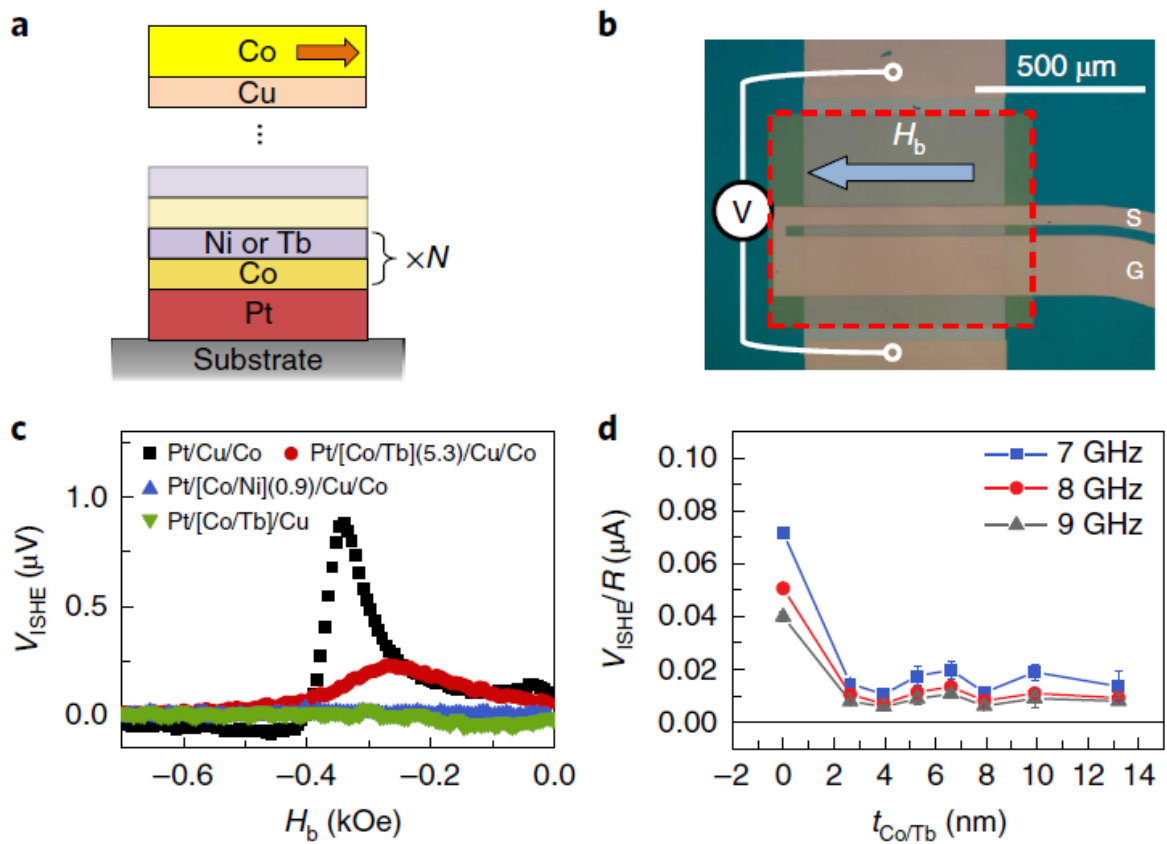

**Figure 16**. Spin pumping measurements in [Co/Tb] and [Co/Ni] multilayer systems. (a) Sample structure for spin pumping measurements. N represents the number of repetition layers. (b) Schematics of the set up for spin pumping measurement. (c) Inverse spin Hall voltage ($V_{ISHE}$) as a function of magnetic field for various sample structures. (d) $V_{ISHE}/R$ as a function of [Co/Tb] thickness ($t_{Co/Tb}$) in Pt/[Co/Tb]/Cu/Co structures at different frequencies. $R$ is the channel resistance. Reproduced with permission from [121] © 2019, Springer Nature.

All these results point to ferrimagnets having tremendous potential and offering large perspectives to be yet fully explored. These materials can be investigated for various purposes and can have applications in many subfields such as quantum information processing, neuromorphic computing, and racetrack memories (the latter rely on the spin of electrons to move data at hundreds of miles per hour to atomically precise positions along a nanowire and may represent the future of data storage).

### 3.5 Perpendicular magnetic anisotropy (PMA) materials

Perpendicular magnetic anisotropy (PMA) materials and interfaces triggered a huge interest after their experimental observation in multilayer thin films [123–125]. PMA plays a very important role in magnetic storage hard disk drive to downscale storage cell size and to provide better thermal stability [126–129]. Modern data storage devices based on magnetic random-access memory (MRAM) require strong perpendicular magnetic anisotropy for ultra-high-density storage and reduction in energy consumption. In addition to this, PMA materials are also required in various magnetism and spintronics research such as spin transistor and topological superconductor for the injection and detection of spins perpendicular to the sample plane [130–132].

Large magnetic anisotropy is usually observed in materials with strong spin-orbit coupling such as heavy elements (rare earths, Bi, Ta, W, Pd, Pt etc.) as the strength of SOC is proportional to $\alpha^2 Z^4$, where $\alpha$ is fine structure constant ($\approx 0.007297$) and $Z$ is the atomic number [12]. However, there are some magnetic metal/ oxide interfaces displaying strong anisotropy despite having weak SOC [12]. In multilayers, the magnetic anisotropy can originate from various effects; (i) broken symmetry at the interfaces can create magnetic anisotropy, (ii) hybridization of electron orbitals at the interfaces produces interfacial anisotropy, (iii) lattice mismatch between structures introduces strain, which also generates anisotropy due to magnetostriction effect, and (iv) the shape of the magnetic specimen due to magnetostatic effect also contribute to magnetic anisotropy [12,133].

The effective anisotropy per unit volume for a magnetic thin film sandwiched between two interfaces can be described as [12]

$$K_{eff} = \left(K_v - \frac{\mu_0}{2}M_s^2\right) + \frac{K_s}{t}$$

S. Gupta, Handbook on the Physics and Chemistry of Rare Earths, Vol. 63, (2023).

Where $\frac{\mu_0}{2}M_s^2$ is the demagnetization energy, $K_v$ represents the bulk anisotropy of the material and $K_s$ combines all the interfacial contributions from the interfaces.

Rare earth iron garnets (REIGs) are promising candidates for spin torque, spin waves, data storage and logic devices as these materials show tunable saturation magnetization, Gilbert damping constant, magnetocrystalline anisotropy, and magnetostriction along with perpendicular magnetic anisotropy properties [134–138]. These garnets are widely used in spintronic research because of their faster magnetization dynamics (in THz regime) and low ohmic losses from parasitic current shunting [138]. Manipulation of magnetization in PMA REIG has been realized in various experiments using adjacent heavy metals [139,140] or topological insulators [141,142] which have a large spin Hall effect.

Bauer *et al.* [138] deposited DyIG thin films on various substrates such as gadolinium gallium garnets (GGG), substituted gadolinium gallium garnets (SGGG) and gadolinium scandium gallium garnets (GSGG) to modify the epitaxial lattice mismatch between the substrate and the deposited structure and thus for tuning the magnetic anisotropy. Figure 17(a) shows the high-resolution X-ray diffraction (HRXRD) pattern of DyIG containing the (444) reflection of single crystalline DyIG along with the (111) reflection from GGG, SGGG and GSGG substrates, labelled as F and S, respectively. Figure 17(b) displays the magnetization data for all the samples carried out using a vibrating sample magnetometer (VSM). It can be noted that the film grown on the GGG substrate has the smallest coercivity among all the measured samples. With the exception of the film grown on the GGG substrate, all samples show perpendicular magnetic anisotropy [138].

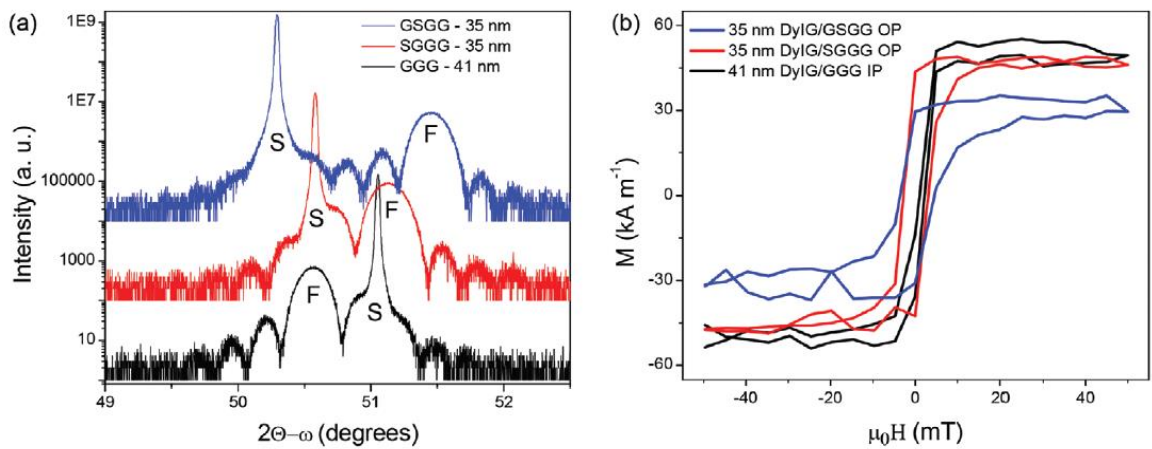

**Figure 17.** (a) High resolution X-ray diffraction pattern of DyIG deposited on different substrates. The labelling F and S stands for film and substrate, respectively. (b) Field dependence of magnetization measured for DyIG on different substrates. Reproduced with permission from [138] © 2020 Wiley-VCH Verlag GmbH.

Perpendicular magnetic anisotropy has also been reported in amorphous PrTMB (TM= Fe, Co) [143] and GdFeCo thin films [144]. The possible origin of PMA in PrTmB has been hypothesized to be residual stress while it was considered to be of structural origin consistent with atomic scale anisotropy in GdFeCo films [143,144]. Figure 18 shows the field dependence of the magnetization of $Gd_{27}Fe_{66}Co_7$ and $Gd_{22}Fe_{71}Co_7$ films as a function of temperature and the room temperature magnetization measured in in-plane and out-of-plane configurations. A clear PMA near $T_M$ (~350 K) can be seen in an as-deposited $Gd_{27}Fe_{66}Co_7$ film (Figure 18a), which can be tuned by varying the concentration of Gd and Fe elements, whereas at other temperatures far away from $T_M$ the film has in-plane magnetization easy axis [144]. Figure 18(b) shows in-plane and out-of-plane magnetization for 50 nm as-deposited $Gd_{22}Fe_{71}Co_7$ film at room temperature, clearly exhibiting PMA. This confirms that by tuning Gd and Fe concentration, anisotropy in these films can be tuned from in-plane to out-of-plane.

S. Gupta, Handbook on the Physics and Chemistry of Rare Earths, Vol. 63, (2023).

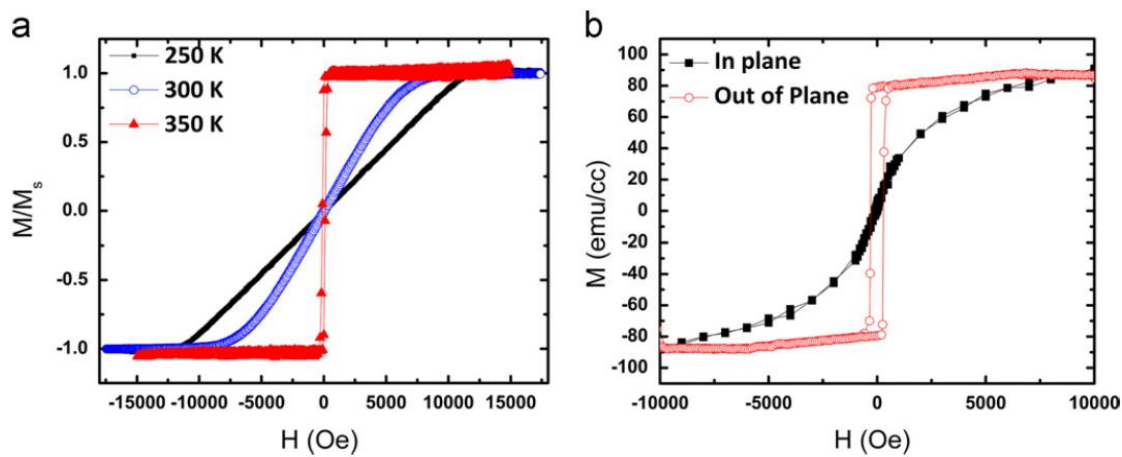

**Figure 18.** (a) The field dependence of magnetization, *M* as a function of temperature for a Gd$_{27}$Fe$_{66}$Co$_7$ film. (b) the field dependence of magnetization, *M* measured in in-plane and out-of-plane configurations at room temperature. Reproduced with permission from [144] © 2013, Elsevier B.V.

### 3.6 Magnetic skyrmion

Magnetic skyrmions are topologically stable swirling magnetic textures, which are a hot research topic nowadays as they present novel characteristics for use in data storage and other spintronic applications. Some rare earth materials such as DyCo$_3$ [145], Gd nano islands [146], CoGd films [116] and GdFeCo were reported to exhibit magnetic skyrmions. Ferromagnetic skyrmions [147,148] have been reported to show large skyrmion Hall angles (>30˚), which is unfavorable for applications such as skyrmion-based racetrack memories [149]. The great tunability of ferrimagnetic rare-earth materials can be very advantageous for the dynamics of magnetic skyrmions as they show negligible stray fields, which can result in small skyrmion Hall angle, which is desirable to realize skyrmion-based racetrack memories [149]. The room temperature stable skyrmion were predicted and later experimentally realized in GdCo film, which exhibit low stray fields along with the bulk PMA that allows skyrmions up to 10 nm size, imaged via element-resolved X-ray holography [116]. A low skyrmion Hall angle (~20˚) has been found in ferrimagnetic GdFeCo (Figure 19a), opening new perspectives for

S. Gupta, Handbook on the Physics and Chemistry of Rare Earths, Vol. 63, (2023).

ferrimagnetic skyrmionics [150]. It has been observed that skyrmons in Gd and FeCo sublayers are antiferromagnetically coupled and can move with a speed of ~50 m/s. Later, Hirata *et al.* [151] reported a vanishing skyrmion Hall effect (Figure 19b) by measuring the elongation angle of a magnetic bubble in GdFeCo at its angular momentum compensation temperature ($T_A$).

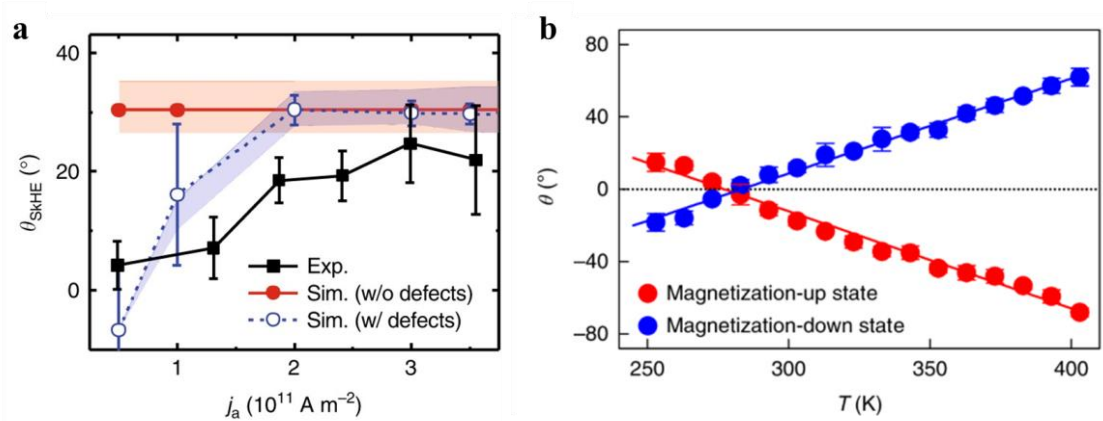

**Figure 19.** (a) Current dependence of skyrmion Hall angle of spin-orbit torque driven skyrmions in GdFeCo. (b) Half skyrmion Hall angle as a function of temperature measured in current induced elongation of magnetic bubble in GdFeCo. The temperature dependence changes sign for magnetization-up and magnetization-down states at the angular momentum compensation temperature $T_A$. Reproduced with permission from: (a) [150] © 2018, Springer Nature, (b) [151] © 2019, Springer Nature.

### 3.7 Two-dimensional rare earth ferromagnets

After the successful exfoliation of graphene in 2004, researchers have very actively started looking for two dimensional (2D) materials [152]. Since 2D materials are atomically thin, they show many promising properties such as high flexibility, better carrier mobility, tunable bandgap, optical transparency, light-matter interaction, and high surface sensitivity. Therefore, these materials are in very high demand for next generation slim and flexible technological devices. For this reason, a variety of 2D materials have been investigated and engineered for specific properties during the past few years. A two-dimensional magnet (2DM)

S. Gupta, Handbook on the Physics and Chemistry of Rare Earths, Vol. 63, (2023).

was experimentally obtained recently. The first intrinsic magnetic order in monolayer/few-layer materials was experimentally observed in $Cr_2Ge_2Te_6$ [153] and $CrI_3$ [154] in 2017. Since then, various 2D materials were investigated for 2DM, however their development is still at a primary stage. Until now only few 2DM including $Cr_2Ge_2Te_6$ and $CrI_3$, $VSe_2$, $Fe_3GeTe_2$, hematite $Fe_2O_3$ and $CrBr_3$ have been experimentally realized [153–156]. Due to their 4f unpaired electrons, rare earths show excellent magnetism and might introduce a new dimension in 2DM for next generation spintronic devices [157]. Recently some 2D RE compounds were synthesized in layered structural form, which mainly include $RX_2$ (R= Eu, Gd; X= Si, Ge), $EuC_6$, RE-chalcogenides [156]. Bulk antiferromagnetism in some of these materials is converted to 2D ferromagnetism, indicating the presence of 2D magnetism [158]. Tokmachev *et al.* [158] have grown $RSi_2$ (R= Eu, Gd) by directing Gd or Eu metal atoms at Si (111) substrate in molecular beam epitaxy system, forming either multilayers or monolayers of $RSi_2$. The authors studied structural, transport and magnetic properties of bulk (multilayer) and few layer $GdSi_2$ and $EuSi_2$. The magnetic and transport properties of $GdSi_2$ are summarized in Figure 20. The bulk $GdSi_2$ has antiferromagnetic nature below ~50 K as can be seen in Figure 20(b). In the thinner film (2 ML), a clear ferromagnetic ground state is observed (Figure 20(c)). It can be noted that a large variation in the Curie temperature, $T_C$, is observed upon changing the applied magnetic field, indicating 2D ferromagnetism [158]. The 2D ferromagnetic signal is so robust that it is easily detectable with superconducting quantum interference device (SQUID), unlike other reported 2DM. Moreover, 2D FM moment and electrical resistivity show strong dependence on film thickness. The electrical resistivity of 1 ML thick $GdSi_2$ is almost nine orders of magnitude larger than that of the bulk $GdSi_2$. This suggests that due to dimensionality and strong correlation effects, the resistivity in ML thick sample is very different to that of its bulk counterpart.

S. Gupta, Handbook on the Physics and Chemistry of Rare Earths, Vol. 63, (2023).

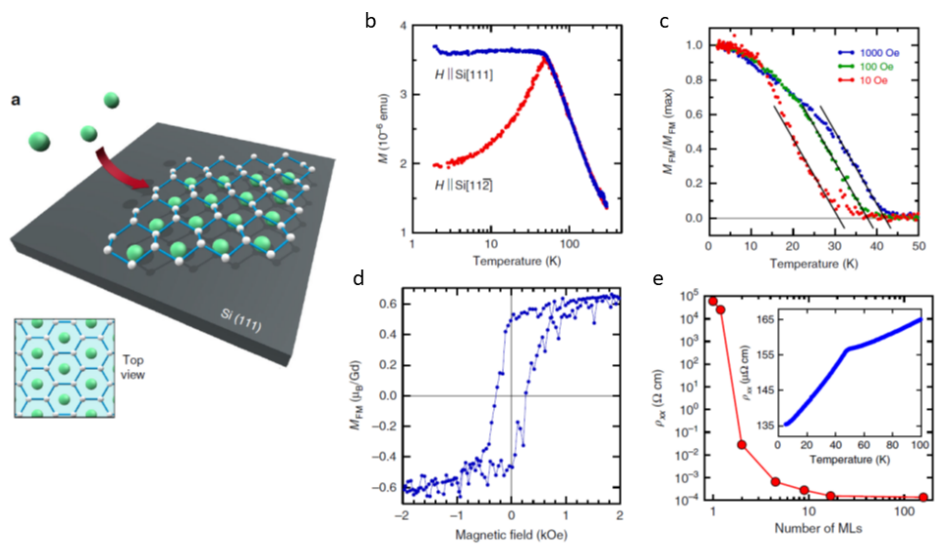

**Figure 20.** (a) Schematic of sample structure. (b) Temperature dependence of bulk GdSi$_2$ (~160 ML) magnetization, showing antiferromagnetic nature. (c) Temperature dependence of normalized magnetization for 2ML thick GdSi$_2$ at different magnetic fields, showing ferromagnetic nature. (d) Magnetic field dependence of magnetization of 2ML thick GdSi$_2$ at 2 K. (e) Thickness dependence of longitudinal resistivity at 2 K. The inset depicts the temperature dependence of the resistivity of bulk GdSi$_2$ (~160 ML). Reproduced with permission from [158] © 2018, Springer Nature

## 4. Conclusion and Outlook

Based on the above discussion, it is obvious that despite of extensive studies on rare earths, the materials described in this review hold promises and need to be further explored. They feature a wealth of properties, ranging from superconductivity to topological magnetism, enabling them as potential candidates for many existing and new technological applications. Strong spin orbit coupling in these compounds results in a variety of novel physical phenomena for spintronics applications. Many rare earths based Heusler alloys, which show semiconducting behavior with very small band gaps, can be investigated theoretically and experimentally for their non-trivial topological states obtained by tuning their electronic structures. The coupling between 4$f$ electrons of rare earths and 3$d$/4$d$ electrons of transition metals make them prime candidates for investigating antiferromagnetic dynamics by tuning angular momenta and magnetization of the magnetic sublattices. The large magnetic moments of rare earths single them out for magnetic topological phase and two-dimensional magnets. It is noteworthy that rare earth materials are presently used to study several new aspects of physics, and that they open new perspectives for novel technologies. Indeed, combining different properties of rare earths and integrating them with other materials leads to intriguing new phenomena.

**Acknowledgement**

The author would like to thank Prof. K. G. Suresh for useful discussions on rare earth materials.

**Abbreviations and acronyms**

| | |
|---|---|
| 2D | Two-dimensional |
| 2DM | Two-dimensional magnet |
| AFM | Antiferromagnetic |
| ARPES | Angle resolved photoemission spectroscopy |
| DFT | Density functional theory |
| DoF | Degree of Freedom |
| DoS | Density of states |
| DW | Domain wall |
| EQHA | Equiatomic quaternary Heusler alloy |
| FCF | Fully compensated ferrimagnet |
| FM | Ferromagnetic |
| GMR | Giant magnetoresistance |
| GGG | Gadolinium gallium garnets |
| GSGG | Gadolinium scandium gallium garnets |
| HA | Heusler Alloy |
| HM | Half metallic |
| HDD | Hard disk drive |
| HRXRD | High-resolution X-ray diffraction |
| ISHE | Inverse spin Hall effect |
| MBE | Molecular beam epitaxy |
| ML | Monolayer |
| MTJ | Magnetic tunnel junction |
| MTI | Magnetic topological insulator |
| MR | Magnetoresistance |
| MRAM | Magnetic random-access memory |
| PM | Paramagnetic |
| PMA | Perpendicular magnetic anisotropic |
| QAHE | Quantum Anomalous Hall effect |
| RE | Rare earth |
| REIG | Rare earth iron garnet |
| SdH | Shubnikov-de Haas |
| SOC | Spin-orbit coupling |
| SOT | Spin orbit torque |
| STT | Spin transfer torque |
| SGGG | Substituted gadolinium gallium garnets |
| SQUID | Superconducting quantum interference device |
| TDS | Topological Dirac semimetal |
| TI | Topological Insulator |
| TIG | Thulium iron garnet |
| TM | Transition metal |
| TMR | Tunnel magnetoresistance |
| TRS | Time reversal symmetry |
| TSS | Topological surface states |
| TWS | Topological Weyl semimetal |
| VSM | Vibrating sample magnetometer |
| YIG | Yttrium iron garnet |

S. Gupta, Handbook on the Physics and Chemistry of Rare Earths, Vol. 63, (2023).

S. Gupta, Handbook on the Physics and Chemistry of Rare Earths, Vol. 63, (2023).

S. Gupta, Handbook on the Physics and Chemistry of Rare Earths, Vol. 63, (2023).

S. Gupta, Handbook on the Physics and Chemistry of Rare Earths, Vol. 63, (2023).